\documentclass[conference]{IEEEtran}
\usepackage{amsmath,amssymb,amsfonts}
\usepackage{algorithmic}
\usepackage{graphicx}
\usepackage{textcomp}
\usepackage{xcolor}
\def\BibTeX{{\rm B\kern-.05em{\sc i\kern-.025em b}\kern-.08em
    T\kern-.1667em\lower.7ex\hbox{E}\kern-.125emX}}
    
\usepackage{svg}   
\usepackage{caption} 
\usepackage{subcaption}
\usepackage[export]{adjustbox} 

\usepackage{hyperref} 
\hypersetup{hidelinks} 

\usepackage{comment}
\usepackage{verbatim}
\usepackage{soul, xcolor}
\usepackage{amsfonts}
\usepackage{booktabs}
\usepackage{multirow}
\usepackage[noadjust]{cite}             
\usepackage[hang,flushmargin]{footmisc} 
\usepackage{textcase}                   
\usepackage[tablename=Table]{caption} 
 \usepackage{epstopdf}

\begin{document}

\title{Subjective Evaluation of Deep Neural Network Based Speech Enhancement Systems in Real-World Conditions
\vspace{-1mm}}

\author{\IEEEauthorblockN{
Gaurav Naithani$^1$, Kirsi Pietil\"a$^2$, Riitta Niemist\"o$^2$, Erkki Paajanen$^2$, Tero Takala$^2$, Tuomas Virtanen$^1$}
\vspace{1mm}
\IEEEauthorblockA{\textit{$^1$Audio Research Group, Tampere University, Tampere, Finland} \\
\small\textit{Email:\{gaurav.naithani, tuomas.virtanen\}@tuni.fi}\\
\vspace{-1mm}\\
\normalsize \textit{$^2$Huawei Tampere Wireless Headset Audio Lab,  Tampere, Finland} \\
\small\textit{Email:\{kirsi.pietila, riitta.niemisto, tero.takala, erkki.paajanen\}@huawei.com}}

\vspace{-7mm}
} 
\maketitle

\begin{abstract}
Subjective evaluation results for two low-latency deep neural networks (DNN) are compared to a matured version of a traditional Wiener-filter based noise suppressor. The target use-case is real-world single-channel speech enhancement applications, e.g., communications. Real-world recordings consisting of additive stationary and non-stationary noise types are included. The evaluation is divided into four outcomes: speech quality, noise transparency, speech intelligibility or listening effort, and noise level w.r.t. speech. It is shown that DNNs improve noise suppression in all conditions in comparison to the traditional Wiener-filter baseline without major degradation in speech quality and noise transparency while maintaining speech intelligibility better than the baseline.
\end{abstract}


\section{Introduction}\label{sec:intro}
Extracting clean speech from noisy speech signals is an important task for various real-world applications \cite{vincent2018audio}, e.g., speech recognition, assisted listening, and mobile communications. The current state-of-the-art methods in the domain of speech enhancement employ some variant of deep neural networks (DNNs), e.g.,  \cite{xu2014regression,lu2013speech, pascual17_interspeech, hu2020dccrn,hao2021fullsubnet}, which have outperformed the classical methods like spectral subtraction \cite{boll1979suppression}, Wiener filtering \cite{ lim1979enhancement} and minimum mean squared error estimation \mbox{(MMSE)  \cite{ephraim1985speech, martin2005speech}}.

Although multi-microphone technologies have gathered a lot of attention over the past decades \cite{alma9910634259105973}, single channel noise reduction remains important as it is robust against variations in microphone locations and device acoustics. In other words, very little tuning is needed in the final stages of production. The main objective of noise suppression is to reduce the background noise level in noisy speech  while preserving speech quality, speech intelligibility, and naturalness of background noise. The old reason to do this is to reduce listener fatigue: codecs are optimized for speech and especially in low bit-rate modes background noise tends to become unnatural and rather annoying. A newer reason is to enable voice conferences and other such communication use cases, where there are more than two lines open, e.g.,  in a conference call it is desirable to hear everyone's voice but no one's background. In such use cases the background noise can be strongly time-varying, e.g., family, pets, bells, and keyboards have been quite typical noise sources during the covid-19 era.

The real-world environments consist of both stationary as well as non-stationary noise types. The latter has been particularly challenging for traditional enhancement \mbox{methods \cite{cohen2001speech}}. Although DNNs have achieved good progress \cite{alexandre20, xu2014regression, goehring2019using} in tackling this, proper evaluation of these methods is still a significant challenge. Most published methods rely on objective metrics computed on simulated mixtures, improvements in which, may not translate to real-world benefits \cite{Gelderblom}.
However, for real-world applications such as mobile communication, the end goal is better measured by subjective  metrics. Moreover, speech in simulated mixtures does not account for various real-world phenomena, e.g., the Lombard effect \cite{brumm2011evolution}. Due to these real-world constraints, it becomes imperative to conduct subjective evaluations of these  models on real-world recordings. Currently, most studies have relied on subjective evaluation on simulated mixtures (e.g., \cite{reddy2020interspeech, reddy2021icassp, rao2021interspeech}), which is a good starting point for evaluation before going to real recordings. Finally, the most reliable results are obtained in conversational tests that require real-time implementation on hardware. 

DNN models for real-world applications, e.g., mobile telephony, are  deployed on edge devices that are constrained by limited computational capacity. The latency of the system should also be small enough to allow for real-time operation. The former can be expressed in terms  memory footprint of the model, i.e., the number of parameters in the model, and, the number of operations often measured in terms of multiply-accumulate additions (MACs). Reducing the computational complexity of these models is a major motivation behind the current research in the field of  knowledge distillation \cite{kim2021test} and model compression \cite{tan2021towards}. Additionally, these models need to process the audio in an online manner and output processed audio with minimum latency. For frequency-domain processing, the algorithmic latency is limited by the size of the synthesis window used for overlap-add processing. 

In this paper, we present subjective listening tests for two low-latency real-time DNN variants: a unidirectional gated recurrent neural network (GRU) \cite{chung2014empirical} and a causal U-Net \cite{ronneberger2015u,choi2018phase}, and compare them with a traditional Wiener-filter baseline. The focus is on evaluating algorithms, and  not their hardware implementations; therefore the tests are done using pre-recorded real-world material which is processed offline. Moreover, conversational tests are also beyond the scope of this work. The algorithmic latency of the models is 20 ms, which is suitable for communications. The networks are trained on simulated mixtures on data recorded using a device mockup. The listening tests are conducted with unsimulated recordings. The included noise types  are stationary (e.g., car), slow-varying (e.g., traffic noise), and non-stationary (e.g., caf\'e) noises.  The separated speech signals are  evaluated by eleven  expert listeners on the following four outcomes: \textit{speech intelligibility}, \textit{speech quality}, \textit{noise transparency}, and \textit{noise level with respect to speech}. Each outcome is rated as an opinion score on a scale of 1 to 5 and the final ratings are then obtained as an average, i.e., mean opinion score (MOS). The subjective evaluation framework used here is based on the ITU-T P.800 \cite{itu-tp800} and ITU-T P.835 \cite{itu-tp835} standards but refined to arrive at four highly perceptual and atomic evaluation attributes.

We show that DNNs are able to improve noise suppression for all noise conditions without considerable degradation in \textit{noise transparency} or \textit{speech quality}, and without any degradation in \textit{speech intelligibility}. Among the two DNN variants, U-Net performed slightly better but has higher computational complexity both in terms of memory footprint measured  as number of model parameters, and, number of computations measured in terms of multiply-accumulate operations (MACs). We also report objective metrics for the two DNN variants on a simulated test set.

\section{Method }\label{sec:method}
\begin{figure}[!t]
\centering
\includegraphics[scale =0.4 ]{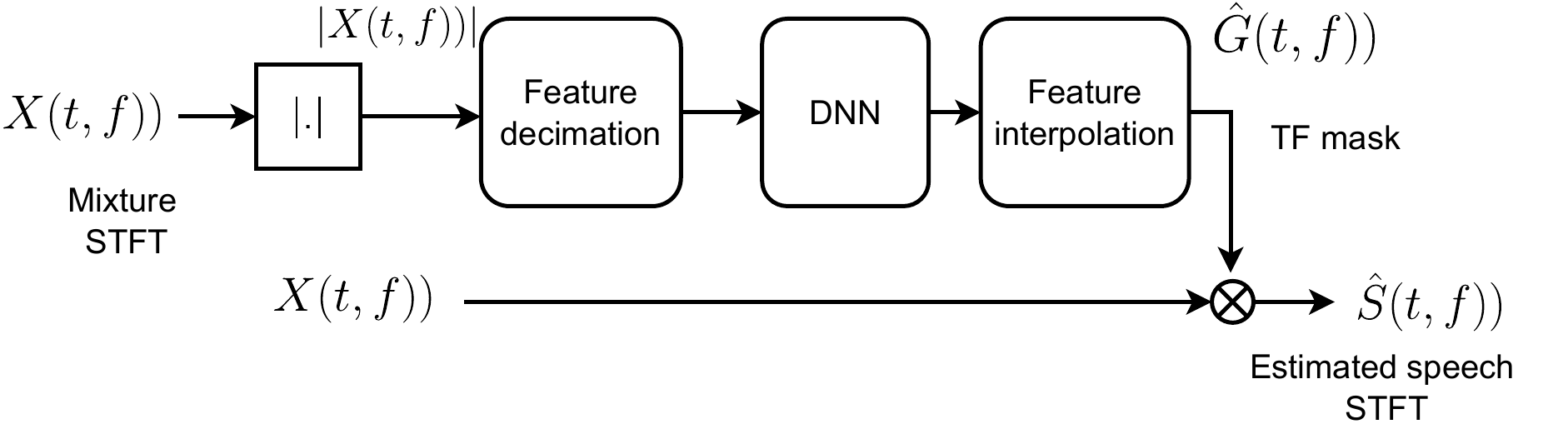}
\caption{\small DNN-based speech enhancement system. }
\label{fig:dnn_block}
\end{figure}

The speech enhancement framework is depicted in  \mbox{Fig. \ref{fig:dnn_block}}. An acoustic mixture is first subjected to  short-time Fourier transform (STFT). Magnitude STFT, $|X(t, f)|$,  is then decimated to a lower resolution for DNN processing. The DNN model is trained to predict a time-frequency (TF)  mask which is then interpolated to restore the original STFT resolution. This TF mask, $\hat{G}(t, f)$,  is then multiplied with mixture STFT to give TF spectra, $\hat{S}(t, f)$,  of the enhanced speech signal. The loss function for DNN training  is  the mean-squared error between the estimated and ground truth speech TF representations. Thus, the mask prediction here is implicit in the sense that the network is trained to optimize directly for the speech TF output which is then subjected to inverse STFT and overlap-add processing to yield the time domain speech.

\subsection{Feature extraction}
The magnitude STFT spectra is computed  on the principle of asymmetric windowing, i.e., the analysis window is longer than synthesis window to allow for better TF resolution,  as was proposed in \cite{mauler2007low, wang_asym}. Since most of the spectral energy for human speech is concentrated in lower frequency bands, and  the human ear is more sensitive to lower frequencies, the higher spectrum can be compressed. This also reduces the input/output size of the DNN and hence computational complexity. Roughly following the perceptually motivated Mel frequency scale, the first 54 of our 513 original frequency bins (FFT length of 1024) are used as such, while the rest of the spectrum is compressed to 12 bands of an increasing width by averaging the spectral bins within each band. The TF mask estimated by DNN is expanded to the original FFT resolution by piecewise constant interpolation.

\begin{figure}[!b]
\centering
\includegraphics[scale=0.4]{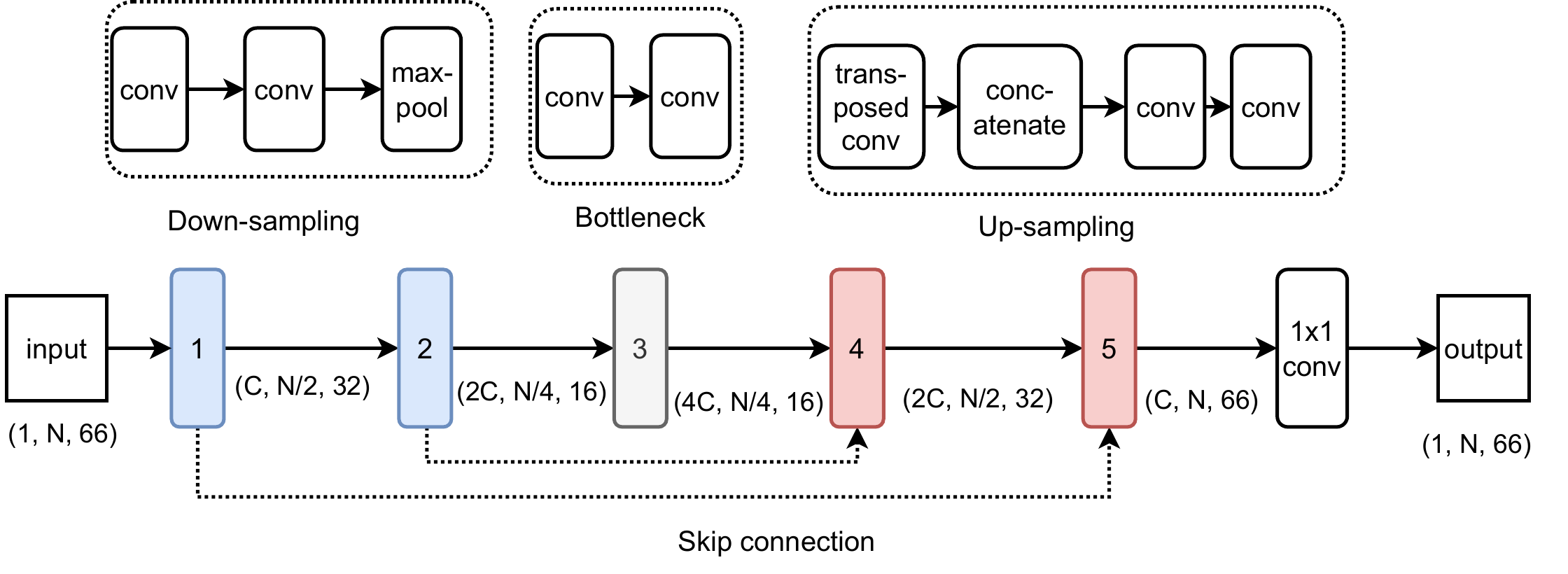}
\caption{\small The U-Net architecture consisting of down-sampling blocks (1, 2),  bottleneck block (3), and up-sampling blocks (4, 5). The sequence of operations in each block is also depicted.}
\label{fig:u-net}
\end{figure}

\subsection{DNN topologies}

Two types of  DNN topologies are used: a unidirectional gated recurrent unit (GRU) network consisting of recurrent layers followed by a feedforward layer, and a U-Net architecture \cite{ronneberger2015u,choi2018phase} consisting of causal convolutional layers. The motivation behind this choice is to test two widely used architectures:   recurrent layers \cite{hao2021fullsubnet, xu14, goehring2019using}, and convolutional layers \cite{hu2020dccrn, bulut, chandna2017monoaural, naithani2017low}, for the real-world speech enhancement task. The U-Net architecture, depicted in Fig. \ref{fig:u-net}, is a sequence of down-sampling and up-sampling blocks with a bottleneck block in between. The network consists of two down-sampling and two up-sampling blocks. Each of these blocks consists of two convolutional layers. There is skip connection between down-sampling  and up-sampling blocks as depicted. The down-sampling is achieved by max-pooling and  up-sampling is achieved through transposed convolutions. The bottleneck block does not have any down/up sampling operation. The sequence of operations is also depicted. The order of operation in an up-sampling block is as follows: the input  is first up-sampled and then concatenated with the skip connection input along the channel dimension. This is then fed to convolutional layers in the block. The network is thus symmetrical. The Fig. \ref{fig:u-net} also shows the output dimension at each stage of processing for an input of shape $(1, \, N,\,  66)$, where $N$ is the number of STFT frames in the input. If $C$ is the number of filters in the convolutional layers in the first down-sampling block, the number of channels at the output of each block is given by,
$1 \textrm{(input)}\rightarrow C\rightarrow 2C\rightarrow 4C\rightarrow 2C\rightarrow C\rightarrow 1 \textrm{(output)},
$
\noindent where the arrows show the processing chain in Fig. \ref{fig:u-net}. The kernel size of  $5 \times 5$ in time and frequency axes is used. The output block is a pointwise $1 \times1 $ convolution block. The convolutional layers are implemented using depth-wise separable convolutions \cite{chollet2017xception} to reduce the computational complexity of the network.

\subsection{Baseline}

The classical Wiener filter, 
$$
    G(t, f) = \frac{\xi(t, f)}{\xi(t, f)+1},\ \ \ \xi(t, f) = \frac{|S(t, f)|^2}{|N(t, f)|^2}
$$
is selected as the ground truth mask that DNNs estimate from input mixture spectrum, $X(t, f) = S(t, f)+N(t, f)$, where $X(t, f)$, $S(t, f)$ and $N(t, f)$ 
refer to STFT of noisy speech, speech and noise, respectively, and $\xi(t, f)$ is the \mbox{a priori} SNR.
Thus, it is fair to compare outputs of DNNs also to a traditional noise suppressor where noise is estimated separately. 
Noise is estimated via
$$
    |\hat{N}(t, f)|^2 = \alpha |\hat{N}(t-1, f)|^2 + \left(1- \alpha\right) |X(t, f)|^2,
$$
and  \mbox{a priori SNR} $\xi(t, f)$ is estimated using "decision directed" approach of \cite{EphraimMalah}  as
$$
    \hat{\xi}(t, f) = \alpha G(t-1, f)^2\gamma(t, f) + (1-\alpha) \max(\gamma(t, f) - 1, 0),
$$
where  \mbox{a posteriori} SNR, $\gamma(t, f) = \frac{ |X(t, f)|^2}{|N(t, f)|^2}$.
The parameter $\alpha$ controls update speeds of the estimates. In our version we have set it separately for each frequency band
and time instant via trial and error using estimates of pitch, other features typical to speech, the spectral distance between noisy speech and noise estimate 
and minimum statistics estimate \cite{Martin94spectralsubtraction} \cite{MartinR2001Npsd}. 
In practical applications \mbox{a priori} SNR is limited below and the limit dictates maximum noise attenuation, typically the limit is set corresponding  to a maximum attenuation of  12 dB.

\begin{table*}[!ht]
\centering
\caption {\small The subjective evaluation framework to assign opinion scores.}
\label{tab:eval}
\vfill
\begin{tabular}{@{}cllll@{}}
\toprule
Score & \textit{Speech intelligibility}   & \textit{Speech quality} & \textit{Noise transparency}                                                                                        & \textit{Noise level w.r.t speech}                                                                                      \\ \midrule
5     & Fully intelligible                          & Excellent      & Natural                                                                                                   & \begin{tabular}[c]{@{}l@{}}Noise level is low compared to speech \\ but not fully muted.\end{tabular} \\
4     & Slightly raised effort needed to understand & Good           & Natural with minor issues                                                                                 & \begin{tabular}[c]{@{}l@{}}Noise level is moderately low, not \\ fully muted\end{tabular}             \\
3     & Clearly raised effort needed to understand  & Fair           & \begin{tabular}[c]{@{}l@{}}Mostly natural with some annoying \\ features or fully muted\end{tabular}      & Noise level is moderate or fully muted                                                                \\
2     & Difficult to understand                     & Poor           & \begin{tabular}[c]{@{}l@{}}More unnatural than natural, e.g., \\ considerable discontinuties\end{tabular} & Noise level is moderately high                                                                        \\
1     & Impossible to understand                    & Bad            & Unnatural                                                                                                & Noise level is high w.r.t speech                                                                      \\ \bottomrule
\end{tabular}
\end{table*}

\section{Evaluation} \label{sec:evaluation}

\subsection{Dataset}
For the training data, speech recordings were captured in an anechoic chamber, quiet listening rooms and  quiet offices. The recordings consist of short sentences spoken in English or in Finnish. In the anechoic chamber, the speakers were instructed to keep voice levels steady while speech levels  were more natural in the recordings that took place in an office or a listening room. There are 29 speakers, most of them native Finnish, aged 25--60,  in the dataset, consisting of both male and female voice types. The noise recordings correspond to various real-world stationary and non-stationary environments like a train station, orchestra tuning session, wind-tunnel, vacuum cleaner noise, etc. The length of each recording is approximately 40 seconds and they were all recorded with four similar headset mockup devices. This dataset was augmented using CSTR-VCTK corpus for speech and TUT acoustic scenes 2017 dataset for noise. The former consists of  speech read from a newspaper by 110 English speakers. Around 20 minutes of speech is available for each speaker. TUT Acoustic Scenes dataset consists of recordings from 15 acoustic scenes, e.g., bus, cafeteria, car, metro station,  etc., and around 52 min of audio is available for each acoustic scene. To compensate for channel mismatch, we first convolved the speech signals with an impulse response derived from the mockup recordings. The impulse responses were calculated to estimate impulse response between the user's mouth and the microphone of the device and since there were many users, one impulse response was randomly selected for each speech signal.  In addition, the speech signal levels were scaled to values randomly sampled from the histogram of speech signal levels derived from the mockup recordings.  



The real-world recordings for subjective evaluation were captured using two different kinds of mockups and one product prototype in Tampere, Finland, and Shanghai, China, in order to ensure that the devised algorithms work properly with different kinds of devices in the same device category. The mockups were an in-house design with microphones and loudspeaker, but without a processing unit, i.e., they can be used for recording and playback in the early stages of algorithm development. Fifteen speakers were involved. The  recorded samples for the listening test were about 10 seconds long. In total 168 such audio samples, i.e., 28 minutes of audio was included. The speakers, aged 25--50, native Finnish and Chinese, spoke English, Finnish and Chinese. The content of the speech was scripted conversations. Both female and male speakers were included. The signal-to-noise ratio (SNR) was not predefined but users talked naturally and had phone conversations in natural environments, e.g., in Tampere railway station, in a car on a highway from Tampere to Helsinki, or  during lunch time in a canteen.  The estimated average SNR ratio of the recordings estimated by averaging signal powers during speech absence and presence is 7 dB. However we must emphasize that this is estimated with noise level changing due to changes in background (e.g., a car passing by) and also the speaker raising and lowering her/his voice. Thus, the estimation is not fully comparable with the SNRs used in generating simulated mixtures. Following noisy scenarios are included: stationary (e.g., car),  slowly varying  (e.g., train station, street, etc.),  non-stationary noises (e.g., metro area, orchestra noise, etc.) including babble (caf\'e) noise. 

\subsection{Experiments}

The DNNs were trained on simulated mixtures while  the evaluation was done on real-world unsimulated mixtures. For training,  the dataset was split into  training and validation folds consisting of 67 \% and 33 \% of the available data. Chunks of speech and  noise recordings  were randomly sampled and summed. The speech/noise signals were first downsampled to 32 kHz. Prior to the summation, a random signal-to-noise ratio (SNR) was chosen in the range [-5, 5] dB. The noise signal was scaled with respect to the speech thereby preserving the speech levels. The mixture signal was then high-pass filtered using a second-order Butterworth filter with a cut-off frequency at 150 Hz in order to suppress low-frequency components. This was done as some types of noises in the dataset were found to be dominated by low-frequency components that cannot be played back with relatively small loudspeakers typical in telephony.  The asymmetric analysis window of effective length 22.5 ms (window  design \cite{wang_asym}) and 50 \% frame shift was used.  The algorithmic latency of the system, determined by length of the synthesis window, is 20 ms. The maximum noise attenuation was limited to 15 dB when processing for the listening tests, i.e., 3 dB more attenuation is allowed than in the baseline method.

A unidirectional GRU network with 1 layer and 128 GRU units followed by a feedforward layer was used. The U-Net configuration, depicted in Fig. \ref{fig:u-net} with $C$ as 16, was used.  These network topologies were chosen from among several network candidates through grid hyper-parameter search based on acceptable objective performance on a simulated test set and desired computational complexity.  The computational complexity for these model architectures in terms of the number of network parameters and multiply-accumulate operations (MACs) per second of input audio during a single inference pass, is reported in Table \ref{tab:obj_metrics}. For training, the Adam optimizer \cite{kingma2014adam} with default learning rate parameters was used as the optimizer. Early stopping was used to stop the training when the validation error stops decreasing for 20 consecutive epochs. The Librosa \cite{mcfee2015librosa} library was used for audio processing and the PyTorch  \cite{paszke2017automatic} framework was used for training the models.

\subsection{Subjective evaluation}\label{sec:eval_listening}

The subjective evaluation was arranged as a blind listening test and 22 experts with a strong background both in critical listening and audio signal processing, with an age range of 26--60 years and known to have normal hearing within their age cohort. Eleven persons participated without the organizer knowing their identity within the invitation list. Each participant anonymously picked a randomized set of listening samples and conducted the evaluation using their high-quality PC sound interface and headphones. 

The evaluation framework consists of computing mean opinion scores (MOS) on four characteristics of the speech processed via DNN/baseline noise reduction: \textit{speech intelligibility}, \textit{speech quality}, \textit{noise transparency}, and \textit{noise level w.r.t speech}. Here, \textit{speech intelligibility} refers to listening effort, according to a rating scale similar to the listening-effort scale of ITU-800 but slightly modified based on our experience to make judging between the scores even easier. \textit{Speech quality} refers to how pleasant and natural the speech sounds according to the respective scale of ITU-T P.800, regardless of how easy it is to understand the speech. For the assessment of the background noise, we have expanded from the one-dimensional background rating scale of ITU-T P.835 to acquire a more diverse understanding of the impact of speech enhancement processing on noise. Our first noise evaluation attribute is \textit{noise transparency} which refers to the similarity of the residual noise in the enhanced speech to the listener's expectation of the original constituent noise. Our definition of transparency is loosely based on \mbox{ITU-R BS.2399-0.} \cite{itu-rbs2399}. The transparency is high when the residual noise is well recognizable and sounds natural, though optimally quieter after noise suppression. \textit{Noise level w.r.t. speech}, on the other hand, indicates the experienced loudness of residual noise in comparison to the experienced loudness of speech. Comparing the result of this attribute for the processed speech to that for the input signal indicates the experienced suppression of noise by the speech enhancement system. The evaluation framework is described in detail in Table \ref{tab:eval}. 

Note that we intentionally did not include an overall quality attribute in the subjective evaluation. Rather, we selected the four attributes described above that are highly perceptual --- and thus objective --- as well as atomic in nature. In our experience, this increases the reliability and reproducibility of the evaluation results. Also, we found it more meaningful to assess the listening effort rather than "hard" word intelligibility. Word intelligibility tends to decrease only in rather harsh noise conditions, whereas listening effort begins to increase already in less severe conditions, thus the latter is more broadly applicable to evaluation.  A limitation of this approach is that none of these attributes can be directly compared to an instrumental overall quality or word intelligibility prediction metric. The chosen subjective attributes and instrumental metrics rather complement each other.

Although the algorithms were trained with 32 kHz data, the  real-world recordings were captured with a sampling rate of 16 kHz. These were then upsampled to 32 kHz for DNN processing. The motivation behind this is a pragmatic  one, i.e.,  using the same model for wideband and super-wideband applications instead of training separate models for each. It was verified that there is not much deviation between the objective metrics for DNNs trained/tested with 16 kHz data and DNNs trained/tested on 32 kHz data (upsampled from 16 kHz as mentioned above). The same was verified in informal listening tests.  The objective metrics considered here are source to distortion ratio (SDR)\cite{vincent2006performance}, short-time objective intelligibility (STOI) \cite{taal2010short}, and perceptual evaluation of speech quality (PESQ) \cite{rix2001perceptual}.

\subsection{Results and discussion}

\begin{figure}[!t]
\centering
\includegraphics[scale=0.165, center]{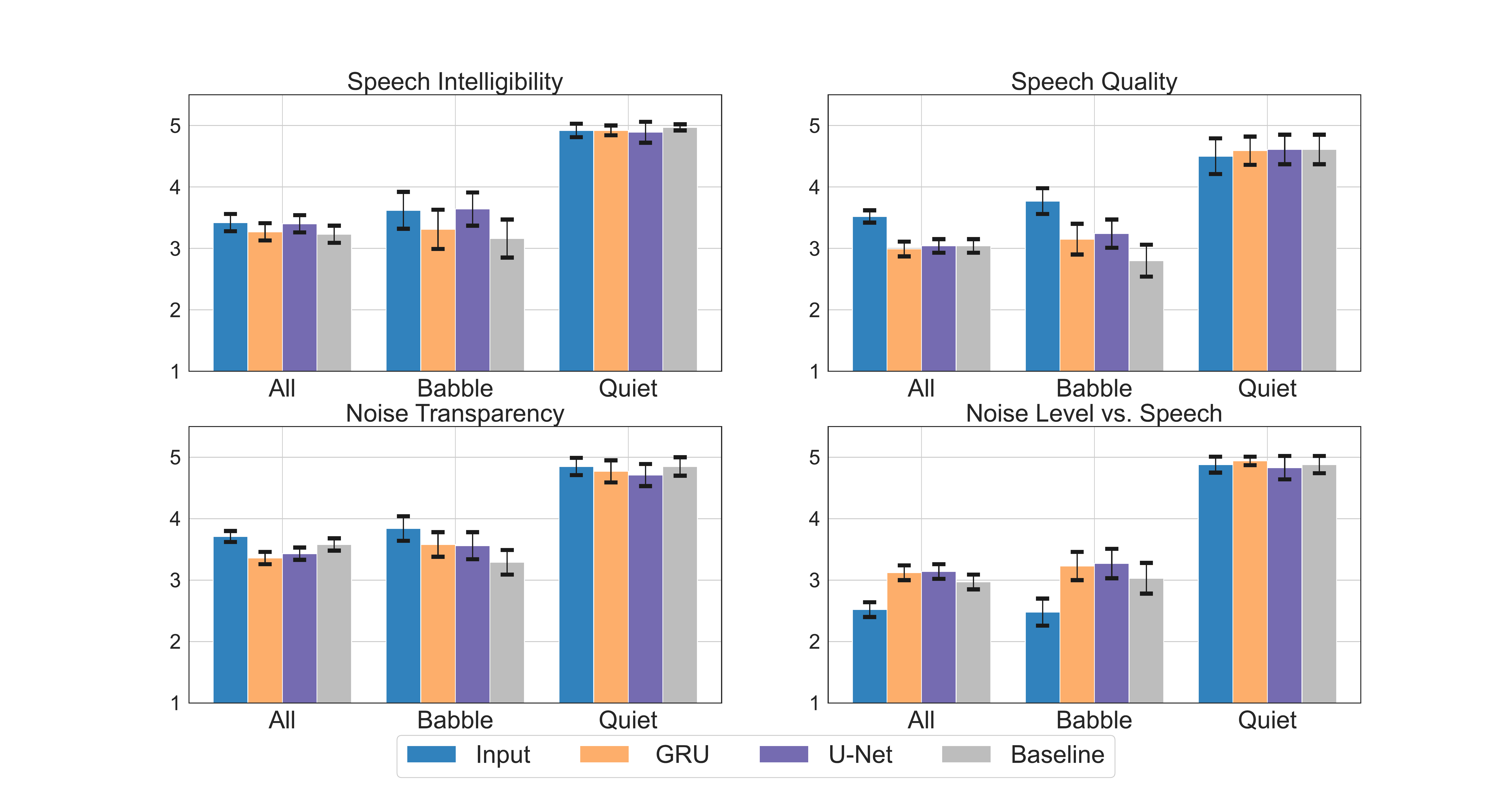}
\caption{\small MOS values for all conditions, babble, and quiet cases separately. The scores are depicted with 80 \% confidence interval.}
\label{fig:agg_noise}
\end{figure}

Fig. \ref{fig:agg_noise} shows the average mean opinion scores averaged over all conditions, babble noise, and quiet cases. We used 80 \% confidence interval in the analysis as we have found it sufficient when comparing the performance of solutions targeted to commercial applications. It is evident that DNNs suppressed noise in all conditions and the 3 dB difference in maximum attenuation explains some, but not all, improvement in speech vs. noise level, namely the difference is higher than average in the babble case. The U-Net preserved \textit{speech intelligibility} slightly better than the GRU, and both DNN variants were better than the baseline. The same was observed in informal listening. Otherwise, the two DNN variants performed similarly with no major statistical difference between them.

We see that, as expected, the use of DNNs was most beneficial in the babble noise case where the noise signal resembles the target signal most. Moreover, we observed that the more stationary the background noise is, the more the listener appreciated the more mature Wiener filtering based baseline solution. Thus, stationary noise cases need more attention, e.g., post-processing or simply more training data. Additionally, on average, DNN processing either preserved \textit{noise transparency} or did not heavily degrade it. The quiet condition in Fig \ref{fig:agg_noise} refers to recording scenarios where practically no background noise or only a low level of background noise was present (quiet offices, listening room, etc.). The DNNs and the baseline method performed equally well in the quiet case.

We also report the objective metrics for the initial and DNN processed conditions corresponding to a simulated test set as well as a comparison of the computational complexity of the solutions in Table \ref{tab:obj_metrics}. A notable observation here is that there is a divergence between  the objective scores on simulated data, i.e., STOI and PESQ, and the subjective scores  on real-world data. Although it is not  fair  to compare the two, it should also be noted that the two modes of evaluation can not be directly compared but rather complement each other as was described in Section \ref{sec:eval_listening}.
 
 Since the baseline method involves a significant amount of smoothing in time and frequency when estimating noise and gain, we expected to have superior performance for DNNs in noise cases where there are sudden changes. In case of stationary or slowly varying noise types no significant improvement is expected. The DNNs at hand also share some limitations of traditional baseline noise suppressor, i.e., time frequency masking corresponds to filtering in the frequency domain with a time-varying zero-phase filter, i.e. it does not modify the phase of the signal and cyclic convolution produces artifacts that are handled in overlap and add processing. Thus, the maximum noise attenuation provided by suppressors is limited so that residual noise masks artifacts. At best noise suppressor preserves the speech and attenuates noise.

\vspace{2mm}

\begin{table}[!t]
\centering
\caption {\small The objective metrics on a simulated test set and computational complexity of the DNNs.}
\label{tab:obj_metrics}
\vfill
\begin{tabular}{@{}lccccc@{}}
\toprule
Model & SDR & STOI & PESQ & Param. & \begin{tabular}[c]{@{}l@{}}MACs \\(per sec) \end{tabular} \\ \midrule
Initial & 0.2   &   0.65  &  2.1  & -        &     - \\
GRU   &  9.7   & 0.67  &   2.5   &     83 K     &  8.5 M     \\
U-Net &  9.6   &  0.68   &  2.6    &     24 K    &  47 M     \\ \bottomrule
\end{tabular}
\end{table}

\section{ Conclusions} \label{sec:conclusion}

In this paper, we evaluated two DNN based speech enhancement systems and reported subjective listening results in the form of average mean opinion scores corresponding to four attributes of separated speech: \textit{speech quality}, \textit{speech intelligibility}, \textit{noise transparency}, and \textit{noise level w.r.t. speech}. We showed that DNNs improve noise suppression in all noisy scenarios without a negative impact on speech intelligibility and without strong degradation of speech quality or noise transparency. Notably, the DNNs are capable of improving speech quality and noise transparency for babble and other non-stationary scenarios, for which traditional, non-DNN based methods are sub-optimal. Out of these selected methods, U-Net seemed to preserve speech intelligibility slightly better than GRU at the expense of higher computational complexity and memory consumption. With these DNNs at least 3 dB more noise attenuation is possible without degrading other quality attributes. For improving other quality attributes besides noise suppression, more sophisticated methods should be considered. Such methods could, e.g., employ multiple microphones or incorporate phase processing.

\section{Acknowledgements}

The authors would like to thank the CSC-IT Centre of Science Ltd., Finland,  for providing computational resources used for the experiments reported in this paper. 

\bibliographystyle{IEEEtran}
\bibliography{mybib}

\end{document}